\documentclass[prl,aps,reprint,floatfix,amssymb,superscriptaddress,longbibliography]{revtex4-2}

\usepackage{amsmath}
\usepackage{amssymb}
\usepackage{bbm}
\usepackage{bm}
\usepackage{graphicx}
\usepackage{mathtools}
\usepackage[dvipsnames]{xcolor}
\usepackage{wasysym}
\usepackage[pdfencoding=auto,naturalnames]{hyperref}
\usepackage{microtype}
\usepackage{times}

\hypersetup{colorlinks=true,citecolor=Blue,urlcolor=Blue,linkcolor=Blue,breaklinks=true}

\newcommand{\ssll}{\hspace{-1em}---}
\newcommand{\Tr}{\operatorname{Tr}}
\newcommand{\ii}{\mathrm{i}}
\newcommand{\avg}[1]{\left\langle #1\right\rangle}

\newcommand{\Ising}{\mathrm{Ising}}

\newcommand{\kket}[1]{\lvert#1\rangle\!\rangle}
\begin{document}

\title{Mixed-State Topological Transition and Emergent Fractionalized Supersymmetry in the Non-Abelian Kitaev Spin Liquid}

\author{Zhou-Quan Wan}
\affiliation{Center for Computational Quantum Physics, Flatiron Institute, New York, New York 10010, USA}

\author{Hong Yao}
\email{yaohong@tsinghua.edu.cn}
\affiliation{Chen-Ning Yang Institute for Advanced Study (YIAS), Tsinghua University, Beijing 100084, China}

\author{Shao-Kai Jian}
\email{sjian@tulane.edu}
\affiliation{Department of Physics and Engineering Physics, Tulane University, New Orleans, Louisiana 70118, USA}

\date{\today}

\begin{abstract}
We study a mixed-state topological transition in the non-Abelian Kitaev spin liquid driven by a bond-dephasing channel.
In the vectorized picture, bond dephasing maps onto interactions between two time-reversed fractionalized Majorana systems coupled to gapped $\mathbb Z_2$ gauge fields, enabling sign-problem-free Monte Carlo simulations.
We show that increasing dephasing condenses an inter-copy fermion pair, converting $\Ising\boxtimes\overline{\Ising}$ to toric-code order in the Choi state and yielding intrinsically mixed-state topological order upon unfolding.
More interestingly, at a temporal interface between topological and trivial Kitaev states, a Chern-number mismatch enforces a gapless Majorana mode. 
Dephasing then drives a mixed-state topological transition at which the gapless mode develops emergent $\mathcal N=1$  supersymmetry.
Our Monte Carlo simulations yield $\nu=1.27(3)$ and $\eta_b\simeq\eta_f\simeq0.40$, consistent with the exact supersymmetric values $5/4$ and $2/5$. The supersymmetric partners are fractionalized and appear physically as distinct gauge-invariant spin strings, providing concrete observables for future experiments. 
\end{abstract}

\maketitle

\paragraph{Introduction.}\ssll
The organizing principles of quantum matter must be enlarged when a system is described by a mixed state rather than a ground-state wave function.
Local dissipative evolution, monitoring, and decoherence motivate notions of phase equivalence that retain information inaccessible to observables linear in the density matrix~\cite{coser2019classification,deGroot2022symmetry,mcginley2019classification,altland2021symmetry,liu2025diagnosing,yang2026topological}.
A particularly sharp example is strong-to-weak spontaneous symmetry breaking (SW-SSB), in which independent left and right symmetry actions of the vectorized density matrix are reduced to their diagonal action~\cite{lee2023quantum,lessa2025strong,sala2024spontaneous,weinstein2025efficient,gu2025spontaneous}.
The resulting order is detected by nonlinear R\'enyi correlators and has now become experimentally accessible in quantum platforms~\cite{wang2026observation,cao2026mixed,wei2026strong,zhang2026probing}.

Decoherence also reorganizes topological order.
Weak noise may leave encoded information recoverable, while stronger noise can drive sharp transitions in coherent information or R\'enyi correlations~\cite{dennis2002topological,bao2023mixed,fan2024diagnostics,sang2023mixed,chen2024symmetry,lee2025symmetry,sala2025decoherence,vadali2026statistical}.
The Choi--Jamio\l kowski map turns this problem into one of ordering in a doubled Hilbert space, where an error channel couples a topological theory to its time-reversed partner~\cite{jamiolkowski1972linear,choi1975completely}.
Anyon condensation in the doubled state can then unfold to ``gauging out'' an anyon of the original density matrix, producing intrinsically mixed-state topological order (imTO) described by nonmodular anyon data and anomalous higher-form symmetry~\cite{wang2025intrinsic,ellison2025toward,sohal2025noisy,zhang2025strong,li2024replica,zhang2026decohering}.
Nevertheless, what universal critical behavior can accompany a transition into such an intrinsically mixed phase remains largely open.

The chiral Kitaev honeycomb spin liquid is a particularly controlled setting for this question.
Its spins fractionalize into itinerant Majorana fermions and static $\mathbb Z_2$ gauge fields, realizing Ising topological order in its non-Abelian phase~\cite{kitaev2006anyons,nayak2008nonabelian}.
The Kitaev model and its associated spin-liquid phases have motivated extensive efforts toward realization in candidate materials and programmable quantum platforms~\cite{jackeli2009mott,hermanns2018physics,broholm2020quantum,evered2025probing,ali2026probing}.
The interplay of topology, fractionalization, and decoherence provides a natural setting for unconventional critical phenomena. 
In particular, interacting fermionic and bosonic degrees of freedom can give rise to emergent spacetime supersymmetry~\cite{friedan1984conformal,lee2007emergence,grover2014emergent,ponte2014emergence,jian2015emergent,rahmani2015emergent,jian2017emergence,li2017edge,ge2025boundary,wu2026emergent}. 
Whether such physics can arise at a decoherence-driven transition involving intrinsically mixed-state topological order remains unexplored.

In this Letter, we uncover unconventional mixed-state criticality in the dephased Kitaev spin liquid. 
We first show that bond dephasing drives inter-copy fermion condensation, converting $\Ising\boxtimes\overline{\Ising}$ into toric-code order in the Choi state; upon unfolding, the maximal-dephasing fixed point realizes imTO with a transparent fermion. 
We then study a temporal interface between topological and trivial Kitaev states, where the Chern-number mismatch enforces a gapless Majorana mode. 
Dephasing tunes this topology-constrained critical phase to a tricritical-Ising endpoint with emergent $\mathcal N=1$ spacetime supersymmetry. 
Our sign-problem-free quantum Monte Carlo simulations yield critical exponents and matching bosonic and fermionic anisotropies, consistent with the supersymmetric predictions. 
We also discuss the supersymmetric partners that are encoded in distinct gauge-invariant spin strings rather than local microscopic operators. 

\begin{figure}[t]
 \centering
 \includegraphics[width=1.0\linewidth]{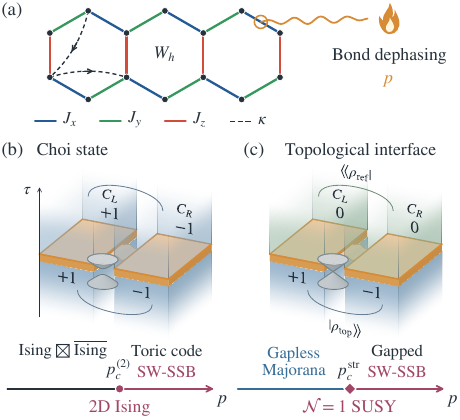}
 \caption{
 (a) Kitaev honeycomb model with bond couplings $J_x,J_y,J_z$ and chiral coupling $\kappa$. Dashed arrows indicate oriented next-nearest-neighbor hopping, and $W_h$ denotes the plaquette flux. 
 Bond dephasing acts as $\mathcal E_{ij}^{\alpha}(\rho)=(1-p)\rho+pB_{ij}^{\alpha}\rho B_{ij}^{\alpha}$.
 (b) Choi-state geometry with no Chern mismatch across the temporal interface. Increasing $p$ condenses the inter-copy boson $\psi\bar\psi$, producing SW-SSB and converting $\mathrm{Ising}\boxtimes\overline{\mathrm{Ising}}$ topological order into toric-code order.
 The transition at $p_c^{(2)}$ belongs to the two-dimensional Ising universality class.
 (c) Temporal interface between topological and trivial states.
 The Chern mismatch $|\Delta C|=1$ supports a gapless Majorana mode at weak dephasing. 
 Increasing $p$ drives a transition into a gapped phase through the fractionalized $\mathcal N=1$ supersymmetric critical point at $p_c^{\rm str}$. 
 }
 \label{fig:schematic}
\end{figure}

\paragraph{Model and doubled Hilbert space.}\ssll
We consider the spin-$1/2$ Kitaev honeycomb model with a time-reversal-breaking three-spin interaction, as shown in Fig.~\ref{fig:schematic}(a),
\begin{equation}
 H= -\sum_{\langle ij\rangle_\alpha}J_\alpha\sigma_i^\alpha\sigma_j^\alpha
 -\kappa\!\sum_{\langle\!\langle ik\rangle\!\rangle}
 \sigma_i^\alpha\sigma_j^\gamma\sigma_k^\beta .
 \label{eq:spin-hamiltonian}
\end{equation}
Here $J_\alpha$ is the exchange coupling on a nearest-neighbor bond $\langle ij\rangle_\alpha$ of type $\alpha=x,y,z$, and $B_{ij}^\alpha\equiv\sigma_i^\alpha\sigma_j^\alpha$ is the corresponding bond operator.
The second sum runs over oriented next-nearest-neighbor pairs $\langle\!\langle ik\rangle\!\rangle$, with $j$ the intermediate site along $i\xrightarrow{\alpha}j\xrightarrow{\beta}k$ and $\gamma$ the remaining spin component.
The $\kappa$ term breaks time-reversal symmetry and can gap the Majorana cones, yielding non-Abelian Ising topological order~\cite{kitaev2006anyons}.

Starting from a ground-state projector $\rho_0$ in this phase, we apply dephasing on each bond~\cite{hwang2024mixed}: 
\begin{equation}
 \mathcal E_{ij}^\alpha(\rho)=(1-p)\rho+p B_{ij}^\alpha\rho B_{ij}^\alpha,
 \qquad 0\leq p\leq\tfrac12,
 \label{eq:channel}
\end{equation}
and the full channel $\mathcal E_p$ is the product over all bonds; we write $\rho_p=\mathcal E_p(\rho_0)$.
This dephasing channel preserves all plaquette fluxes $W_h=\prod_{n=1}^{6}\sigma_n^{\beta_n}$, where the sites are ordered around hexagon $h$ and $\beta_n$ labels the bond leaving the hexagon at site $n$.
The initial ground state lies in the zero-flux sector, with $W_h=+1$ on every hexagon.
Since $[B_{ij}^\alpha,W_h]=0$ for every bond and plaquette, $\rho_p$ remains in this sector.

To study the mixed-state ordering induced by dephasing, we vectorize the density matrix as $\lvert\rho\rangle\!\rangle=(\rho\otimes\mathbbm 1)\lvert\Omega\rangle\!\rangle$, where $\lvert\Omega\rangle\!\rangle=\prod_i(\lvert\uparrow\uparrow\rangle_i+\lvert\downarrow\downarrow\rangle_i)/\sqrt2$ and $O_L\lvert\Omega\rangle\!\rangle=O_R^T\lvert\Omega\rangle\!\rangle$~\cite{jamiolkowski1972linear,choi1975completely}.
The doubled Hilbert space consists of a left copy ($L$) and a right copy ($R$), representing the ket and bra indices of $\rho$, respectively.
Writing $D_{ij}^\alpha=B_{ij,L}^\alpha B_{ij,R}^\alpha$, the channel in Eq.~\eqref{eq:channel} becomes
\begin{equation}
\widehat{\mathcal E}(p)=
\prod_{\alpha,\langle ij\rangle_\alpha}
[(1-p)\mathbbm1+pD_{ij}^\alpha]
\propto e^{V(p)\sum_{\alpha,\langle ij\rangle_\alpha}D_{ij}^\alpha},
\label{eq:doubled-channel}
\end{equation}
where $V(p)=-\tfrac12\ln(1-2p)$.
This exponential representation is exact up to an overall scalar: $(D_{ij}^\alpha)^2=1$, and all such operators commute because anticommutation signs cancel between the two copies.
Thus dephasing generates an inter-copy coupling whose strength increases with $p$.

\paragraph{Fractionalization.}\ssll
We fractionalize each spin as $\sigma_{i,s}^\alpha=\ii b_{i,s}^\alpha c_{i,s}$, subject to the local constraint $G_{i,s}=b_{i,s}^xb_{i,s}^yb_{i,s}^zc_{i,s}=1$, where $s=L,R$.
Following the convention detailed in the Supplemental Material~\cite{supplemental}, we define $u_{ij,s}=\ii\zeta_s b_{i,s}^\alpha b_{j,s}^\alpha$, with $\zeta_L=-\zeta_R=1$ and $u_{ji,s}=-u_{ij,s}$.
The physical state is obtained by gauge projection, while $W_{h,s}=\prod_{\langle ij\rangle\in\partial h}u_{ij,s}$ is gauge invariant; the plaquette product uses the fixed $A$-to-$B$ orientation on every bond. 
The bond-dephasing channel commutes with all static link operators $u_{ij,s}$ and hence preserves each flux sector. 
We work in a fixed zero-flux sector with $u_{ij,L}=u_{ij,R}$; the boundary conditions and physical projection are specified in the Supplemental Material~\cite{supplemental}.
The matter Hamiltonians then read
\begin{equation}
 H_s=\zeta_s\left(\ii\!\sum_{\alpha,\langle ij\rangle_\alpha}J_\alpha c_{i,s}c_{j,s}+\ii\kappa\!\sum_{\langle\!\langle ik\rangle\!\rangle}
 \nu_{ik}c_{i,s}c_{k,s}\right).
\label{eq:majorana-hamiltonian}
\end{equation}
Here $\nu_{ik}=\epsilon_{\alpha\beta\gamma}=-\nu_{ki}$, with $\epsilon_{xyz}=1$, for the oriented path defined above.
The two Hamiltonians satisfy $H_R=H_L^*$ and therefore have opposite matter-band Chern numbers.

In this gauge convention, the bond-dephasing channel in Eq.~\eqref{eq:doubled-channel} becomes
$\widehat{\mathcal E}(p)\propto\exp[V(p)\sum_{\langle ij\rangle}\Phi_i\Phi_j]$, corresponding to a ferromagnetic coupling of the inter-copy Majorana bilinear $\Phi_i=\ii c_{i,L}c_{i,R}$.
This provides the tunable inter-copy interaction that will later control the temporal-interface criticality.
An exact Hubbard--Stratonovich decomposition and Majorana positivity make this interacting problem accessible to sign-problem-free Monte Carlo simulations; the construction and estimators are given in the Supplemental Material~\cite{supplemental,li2015solving,li2016majorana,wei2016majorana,YuanPRB2026}.

\paragraph{Topological transition.}\ssll
We first consider the normalized Choi state, whose overlap geometry is illustrated in Fig.~\ref{fig:schematic}(b).
The coupling $V(p)\sum_{\langle ij\rangle}\Phi_i\Phi_j$ grows monotonically with dephasing and eventually drives an inter-copy pairing instability.
However, $\Phi_i$ is gauge charged: under the independent $\mathbb Z_2$ gauge transformations of the two copies, $c_{i,s}\to\lambda_{i,s}c_{i,s}$ and $u_{ij,s}\to\lambda_{i,s}u_{ij,s}\lambda_{j,s}$, it transforms as $\Phi_i\to\lambda_{i,L}\lambda_{i,R}\Phi_i$.
A physical pairing correlator must therefore connect its endpoints by a $\mathbb Z_2$ gauge string in each copy.

For an oriented path $\gamma:i\to j$, define the Wilson string $U_{\gamma,s}=\prod_{\langle lm\rangle\in\gamma}u_{lm,s}$. 
The gauge-invariant R\'enyi-2 pairing correlator of the Choi state $\lvert\rho_p\rangle\!\rangle=\widehat{\mathcal E}(p)\lvert\rho_0\rangle\!\rangle$ is
\begin{equation}
    \mathcal C_\Phi^{(2)}(i,j;\gamma)= \frac{\langle\!\langle\rho_p|\Phi_iU_{\gamma,L}U_{\gamma,R}\Phi_j|\rho_p\rangle\!\rangle}{\langle\!\langle\rho_p|\rho_p\rangle\!\rangle}.
\label{eq:gauge-invariant-boson}
\end{equation}
The Wilson strings cancel the endpoint gauge charges.
Since the two copies have identical link configurations, $U_{\gamma,L}U_{\gamma,R}=1$, so this expression reduces to the normalized Choi correlator of $\Phi_i\Phi_j$.
It is unchanged under contractible path deformations within the zero-flux sector.

\begin{figure}[t]
 \centering
 \includegraphics[width=1.0\linewidth]{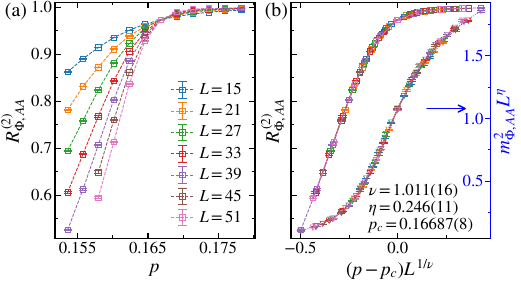}
 \caption{Inter-copy condensation in the normalized Choi state.
 (a) R\'enyi-2 Wilson-string correlation ratio $R_\Phi^{(2)}=1-S_\Phi^{(2)}(\bm q_{\min})/S_\Phi^{(2)}(\bm0)$ for $L=15,21,\ldots,51$. 
 Here $S_\Phi^{(2)}(\bm q)=L^{-2}\sum_{i,j\in A} e^{\ii\bm q\cdot(\bm r_i-\bm r_j)}\mathcal C_\Phi^{(2)}(i,j;\gamma_{ij})$ and $\bm q_{\min}$ is the smallest nonzero momentum.
 (b) Finite-size collapses of the correlation ratio (left axis) and $m_\Phi^2L^\eta$ (right axis), with $m_\Phi^2=S_\Phi^{(2)}(\bm0)/L^2$, versus $(p-p_c^{(2)})L^{1/\nu}$.
 The fits yield $p_c^{(2)}=0.16687(8)$, $\nu=1.011(16)$, and $\eta=0.246(11)$, consistent with the two-dimensional Ising values $\nu=1$ and $\eta=1/4$. 
 See Supplemental Material~\cite{supplemental} for details.}
 \label{fig:top-transition}
\end{figure}

For $J_x=J_y=J_z=1$ and $\kappa=1/2$, we take $\lvert\rho_0\rangle\!\rangle$ from the non-Abelian Ising topological phase.
Fig.~\ref{fig:top-transition}(a) shows a clear common crossing of the dimensionless correlation ratio $R_\Phi^{(2)}$.
Its finite-size flow reverses across the crossing, decreasing with $L$ on the weak-dephasing side and increasing on the strong-dephasing side, signaling a transition from the disordered phase to an inter-copy condensed phase.
A finite-size collapse of $R_\Phi^{(2)}$ and the pairing order parameter $m_\Phi^2$ in Fig.~\ref{fig:top-transition}(b) yields $p_c^{(2)}=0.16687(8)$, $\nu=1.011(16)$, and $\eta=0.246(11)$, consistent with the two-dimensional Ising transition, $\nu=1$ and $\eta=1/4$.

The ordered phase identified above has a direct symmetry interpretation. 
Since $\Phi$ is odd under either copy fermion parity but even under their product, its long-range order realizes the fermion-parity SW-SSB pattern $\mathbb Z_{2,L}^{f}\times\mathbb Z_{2,R}^{f}\to\mathbb Z_{2,\mathrm{diag}}^{f}$.
From the anyon perspective, after gauge projection a matter-Majorana endpoint corresponds to the Ising fermion anyon $\psi$ and must be attached to a $\mathbb Z_2$ Wilson line. 
The gauge-invariant string operator in Eq.~\eqref{eq:gauge-invariant-boson} therefore creates a $\psi\bar\psi$ boson at each endpoint.
Thus the transition is an anyon $\psi\bar\psi$ condensation transition in the normalized Choi state~\cite{li2024replica}, as shown in Fig.~\ref{fig:schematic}(b).

Viewed as a pure state in the doubled Hilbert space, $\lvert\rho_0\rangle\!\rangle$ has $\Ising\boxtimes\overline{\Ising}$ topological order, with sectors $a\bar b$ for $a,b\in\{1,\sigma,\psi\}$.
Condensation of $\psi\bar\psi$ identifies sectors related by fusion
with the condensate and confines those with nontrivial monodromy.
Thus $1\bar1\sim\psi\bar\psi$ and $\psi\bar1\sim1\bar\psi\equiv\epsilon$.
The sector $\sigma\bar\sigma$ is invariant under fusion with $\psi\bar\psi$ and, having quantum dimension two, splits as $e+m$.
Consequently~\cite{bais2009condensate,kong2014anyon,burnell2018anyon,hwang2024anyon},
\[
 \Ising\boxtimes\overline{\Ising}
 \xrightarrow{\;\psi\bar\psi\;}
 \{1,e,m,\epsilon\}_{\rm TC},
\]
yielding the toric-code topological order.

Unfolding changes the interpretation of this condensation.
The right copy is the bra index, and $\psi\bar\psi$ represents the quantum channel $\rho\mapsto W_\psi\rho W_\psi^\dagger$ rather than an independent bilayer anyon.
The transition therefore gauges out $\psi$ instead of simply declaring it the vacuum~\cite{sohal2025noisy}.
This gauging-out process has a higher-form SW-SSB manifestation.
Because a one-sided $\sigma$ Wilson loop has monodromy $-1$ with $\psi\bar\psi$, its independent left and right actions cease to be strong symmetries, while their diagonal action survives as a weak symmetry.
By contrast, the $\psi$ Wilson loop has trivial monodromy with the condensate and remains a strong symmetry.
The surviving strong sector is therefore $\{1,\psi\}$, in which $\psi$ is a transparent fermion, rendering the sector nonmodular.
Assuming no intervening transition, the ordered branch is continuously
connected to the maximal-dephasing endpoint, which is an imTO fixed
point~\cite{ellison2025toward,sohal2025noisy,zhang2025strong}.

\paragraph{Emergent supersymmetry.}\ssll
Having established the bulk dephasing transition and its mixed-state topological interpretation, we now ask how the same dephasing acts on an interface enforced by a change of topology.
We construct the temporal interface shown in Fig.~\ref{fig:schematic}(c) by pairing the non-Abelian Kitaev state with an Abelian reference state whose matter band is topologically trivial.
This interface is naturally probed by the mixed-state strange correlator~\cite{you2014strange,zou2023channeling,lepori2023strange,su2024conformal,sarma2025effective},
\begin{equation}
\begin{aligned}
    \avg{\mathcal O}_{\rm str}&= \frac{ \langle\!\langle\rho_{\rm ref}\rvert\mathcal O\widehat{\mathcal E}(p) \lvert\rho_{\rm top}\rangle\!\rangle}{\langle\!\langle\rho_{\rm ref}\rvert \widehat{\mathcal E}(p)\lvert\rho_{\rm top}\rangle\!\rangle}.
\end{aligned}
\label{eq:strange}
\end{equation}
Here $\lvert\rho_{\rm top}\rangle\!\rangle$ and
$\lvert\rho_{\rm ref}\rangle\!\rangle$ are the vectorized ground-state
projectors of Eq.~\eqref{eq:spin-hamiltonian} with
$J_x=J_y=J>0$, $\kappa>0$, $0<J_z^{\rm top}<2J$, and
$J_z^{\rm ref}>2J$, respectively.
Unlike the normalized Choi correlator in Eq.~\eqref{eq:gauge-invariant-boson}, the strange overlap involves a single factor of $\widehat{\mathcal E}(p)$ and different boundary states.
Its critical probability, denoted by $p_c$, need not coincide with $p_c^{(2)}$.
The Chern-number mismatch produces one chiral Majorana mode in each copy~\cite{read2000paired}, forming a nonchiral Majorana field whose bilinear mass is forbidden by the independent copy parities; the temporal interface is therefore gapless at weak dephasing.
To obtain its propagator, combine the two sublattice matter Majoranas, after a fixed basis rotation, into
$f_{\bm q,s}=(c_{A,\bm q,s}+\ii\zeta_s c_{B,\bm q,s})/2$.
Near the band inversion, the topological state has a singular BCS pair amplitude, while the trivial reference state can be flattened to the empty $f$ vacuum without affecting this singularity.
Time reversal relates the two copies, so at $p=0$,
\begin{equation}
    \langle f_{L,\bm q}f_{L,-\bm q}\rangle_{\rm str}=-\langle f_{R,\bm q}f_{R,-\bm q}\rangle_{\rm str}^{*} \sim \frac{1}{v_{fx} q_x+\ii v_{fy} q_y}.
\label{eq:majorana-propagators}
\end{equation}
The detailed BCS-overlap derivation is given in the Supplemental Material~\cite{supplemental}.
The inverse of the singular propagators in
Eq.~\eqref{eq:majorana-propagators} fixes the universal quadratic action,
\begin{equation}
    S_M[\chi]=\frac12\int\frac{d^2q}{(2\pi)^2}\, \chi^T_{-\bm q}(\ii v_{fy} q_y\mathbbm1+v_{fx} q_x\sigma^z)\chi_{\bm q},
 \label{eq:majorana-action}
\end{equation}
where $\chi$ denotes the long-wavelength chiral Majorana field associated with the singular part of the microscopic $f$ field and $\sigma^z$ acts in the $(\chi_L,\chi_R)$ copy/chirality space. 
The two components are exchanged by the time-reversal-like copy operation and together form a nonchiral Ising Majorana field theory with central charge $c=1/2$.

\begin{figure*}[tb]
 \centering
 \includegraphics[width=1.0\linewidth]{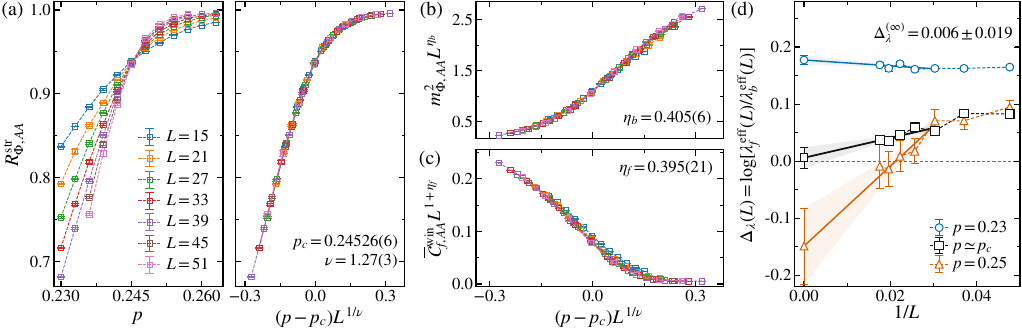}
 \caption{Emergent supersymmetry at the temporal interface. (a) Inter-copy correlation ratio $R_{\Phi,AA}^{\rm str}$ for $L=15,21,\ldots,51$ (left) and its finite-size collapse (right), yielding $p_c=0.24526(6)$ and $\nu=1.27(3)$.
 (b),(c) Data collapses of the rescaled bosonic estimator $m_{\Phi,AA}^2L^{\eta_b}$ and window-averaged fermionic string correlator $\overline{\mathcal C}_{f,AA}^{\rm win}L^{1+\eta_f}$, respectively.
 Independent fits give $\eta_b=0.405(6)$ and $\eta_f=0.395(21)$, consistent with the supersymmetry predictions $\eta_b=\eta_f=2/5$ and $\nu_p=5/4$. The fits in (b),(c) use $p_c$ and $\nu$ from (a), with their uncertainties propagated through bootstrap resampling. 
 (d) Finite-size anisotropy mismatch $\Delta_\lambda(L)=\ln[\lambda_f^{\rm eff}(L)/\lambda_b^{\rm eff}(L)]$ versus $1/L$, with $\lambda_b^{\rm eff}$ and $\lambda_f^{\rm eff}$ extracted from directional $AA$ correlation ratios. Solid lines show linear fits in $1/L$, and symbols at $1/L=0$ denote the extrapolated intercepts. 
 The near-critical data yield an intercept of $0.006(19)$, consistent with zero mismatch, marked by the horizontal dashed line. 
 Details of the simulation setup and finite-size scaling analysis are provided in the Supplemental Material~\cite{supplemental}.
 }
 \label{fig:susy-transition}
\end{figure*}

The gauge sector contributes no additional critical mode: plaquette fluxes are conserved, so the dynamics remains exactly within the zero-flux sector.
More generally, gapped vison excitations only renormalize local couplings and do not introduce additional relevant perturbation~\cite{kitaev2006anyons,wu2026emergent}.
The relevant degrees of freedom at the criticality are therefore the inter-copy scalar $\phi$---the continuum counterpart of the gauge-invariant completion of $\Phi$---and the gapless Majorana mode $\chi$. 
Including the scalar kinetic, mass, and quartic terms gives the Majorana--Ising Yukawa action
\begin{eqnarray}
    S &=& S_M[\chi] + \int d^2x \,\big( \tfrac1 2 [(v_{bx}\partial_x\phi)^2+ (v_{by}\partial_y\phi)^2] \nonumber \\
    && + \tfrac r 2\phi^2 +\tfrac{u}{4!}\phi^4 +g\phi\,\ii\chi_R\chi_L \big),
\label{eq:yukawa}
\end{eqnarray}
where $r$ and $u$ are the quadratic and quartic scalar couplings, and $g$ is the Yukawa coupling.

Since dephasing preserves the copy parities, varying $p$ tunes interactions along the gapless Ising manifold until $\phi$ becomes critical.
Importantly, emergent supersymmetry does not require microscopic spatial isotropy. 
It has been shown that the relative boson--fermion anisotropy, $\lambda_f/\lambda_b$, with $\lambda_a=v_{a y}/v_{a x}$ for $a=b,f$, flows to one, $(\lambda_f/\lambda_b)_\ast = 1$,  at the infrared fixed point~\cite{grover2014emergent,jian2015emergent,wu2026emergent}. 
Thus the bosonic and fermionic metrics flow to a common form, which may be chosen isotropic by coordinate rescaling. 
After this rescaling, the renormalized couplings approach the supersymmetric manifold $r_*=0$ and $u_*=3g_*^2$.
The resulting fixed point is the tricritical-Ising CFT, with $c=7/10$ and emergent $\mathcal N=1$ spacetime supersymmetry~\cite{friedan1984conformal,difrancesco1997cft,ginsparg1988applied,obrien2018lattice}. 
At the fixed point, the bosonic field $\phi$ overlaps with $\epsilon$, with scaling dimension $x_\epsilon=1/5$, while the Majorana field $\chi$ overlaps with its fermionic superpartner $\psi_\epsilon$, with $x_{\psi_\epsilon}=7/10$~\cite{friedan1985superconformal}.
Their correlations therefore decay as $\langle\phi(\bm r)\phi(\bm0)\rangle_{\rm str}\sim r^{-\eta_b}$ and $\langle\chi(\bm r)\chi(\bm0)\rangle_{\rm str}\sim r^{-(1+\eta_f)}$.
Supersymmetry requires $\eta_b=\eta_f$, with the common value $2/5$ at the tricritical-Ising fixed point.

We test these predictions using gauge-invariant lattice string correlators.
The bosonic correlator  uses the Wilson-line completion introduced above, now evaluated with $\mathcal{C}_b(i,j;\gamma)=\langle\Phi_iU_{\gamma,L}U_{\gamma,R}\Phi_j\rangle_{\rm str}$. 
Its fermionic counterpart is $\mathcal{C}_f(i,j;\gamma)=\ii\left\langle c_{i,L}U_{\gamma,L}c_{j,L}\right\rangle_{\rm str}$.
The bosonic and fermionic strings are type-II and type-I strange correlators, respectively~\cite{lee2025symmetry}.
They provide gauge-invariant probes of the fractionalized supermultiplet~\cite{wu2026emergent}.

For finite-size analysis, we use the bosonic diagnostics
$R_\Phi^{\rm str}$ and $m_\Phi^2$, defined analogously
to those in the Choi-state analysis, together with the
window-averaged fermionic correlator
\begin{equation}
  \overline{\mathcal C}_f^{\rm win}
  =\frac{1}{|\mathcal W_L|}
  \sum_{\bm d\in\mathcal W_L}
  \left|\mathcal C_{f,AA}(\bm d)\right|.
  \label{eq:numerical-observables}
\end{equation}
All estimators use translation-averaged $AA$ correlations, and $|\mathcal W_L|$ counts the displacements in the averaging window.
At criticality, $m_\Phi^2\sim L^{-\eta_b}$ and $\overline{\mathcal C}_f^{\rm win}\sim L^{-1-\eta_f}$.
The spin-string representations, lattice and window conventions, and numerical procedures are detailed in the Supplemental Material~\cite{supplemental}. 
Fig.~\ref{fig:susy-transition}(a) shows a common crossing of $R_\Phi^{\rm str}$ for $L=15,21,\ldots,51$.
Its finite-size collapse gives $p_c=0.24526(6)$ and $\nu=1.27(3)$, consistent with the predicted $\nu_p=5/4$~\cite{zamolodchikov1991tricritical}.
Holding $p_c$ and $\nu$ fixed, independent fits of the bosonic and fermionic observables in Fig.~\ref{fig:susy-transition}(b)(c) yield $\eta_b=0.405(6)$ and $\eta_f=0.395(21)$, consistent with the emergent supersymmetry.

Beyond the scaling exponents, we compare the spatial anisotropies of the two sectors using their $AA$ correlators at $\bm d_\pm=\frac{L}{3}(\bm a_1\pm\bm a_2)$.
These displacements lie along the Cartesian axes and satisfy $|\bm d_-|/|\bm d_+|=\sqrt3$.
Motivated by the anisotropic critical power laws, we define the finite-size estimators
\begin{equation}
  \lambda_a^{\rm eff}(L)
  =\sqrt3\left|
  \frac{\mathcal C_{a,AA}(\bm d_-)}
       {\mathcal C_{a,AA}(\bm d_+)}
  \right|^{1/\alpha_a},
  \qquad a=b,f,
  \label{eq:anisotropy-estimator}
\end{equation}
where we fix $\alpha_b=2/5$ and $\alpha_f=7/5$ to the supersymmetric predictions.
Fig.~\ref{fig:susy-transition}(d) shows the relative anisotropy $\Delta_\lambda(L) =\log[\lambda_f^{\rm eff}(L)/\lambda_b^{\rm eff}(L)]$, together with fits to $\Delta_\lambda(L)=\Delta_\lambda^{(\infty)}+A_\lambda/L$.
At $p\simeq p_c$, the extrapolated mismatch is compatible with zero, providing a complementary consistency check of matching bosonic and fermionic correlation anisotropies.

\paragraph{Concluding remarks.}\ssll
We have identified two complementary universal structures generated by bond dephasing of a non-Abelian Kitaev spin liquid.
In the two-dimensional Choi construction, a R\'enyi-2 Wilson-string correlator locates an inter-copy anyon condensation transition from doubled-Ising to toric-code order; unfolding this condensate at the maximal-dephasing fixed point gives imTO with a transparent fermion.
More interestingly, at a temporal interface with a trivial reference state, the Chern mismatch supplies gapless Majorana modes.
The resulting transition is instead the tricritical-Ising fixed point with emergent fractionalized  $\mathcal N=1$ supersymmetry.

Our predictions are in principle accessible on programmable spin platforms that can prepare topological states and probe anyonic observables~\cite{satzinger2021realizing,semeghini2021probing,bluvstein2022quantum}. 
The gauge-invariant bosonic and fermionic strings can be accessed through a spin-based two-copy measurement protocol involving local Bell readout, with the prescribed dephasing incorporated in classical postprocessing.
Beyond the integrable limit, we expect this criticality to persist while the vison gap remains open and gauge fluctuations generate no additional relevant perturbation.

{\it Acknowledgments:} 
This work is supported in part by the NSFC under Grant Nos. 12347107 and 12334003 (H.Y.)
and by the New Cornerstone Science Foundation through the Xplorer Prize (H.Y.).
The Flatiron Institute is a division of the Simons Foundation. We used the AI tool GPT models (GPT-5.6 and GPT-6) to assist with manuscript editing.

\makeatletter
\renewcommand{\bibfont}{\small\linespread{0.96}\selectfont\@clubpenalty\clubpenalty}
\makeatother
\bibliography{references}

\clearpage

\setcounter{section}{0}
\setcounter{subsection}{0}
\setcounter{secnumdepth}{2}
\setcounter{tocdepth}{2}

\begin{widetext}

\begin{center}
{\large\bf Supplemental Material for ``Mixed-State Topological Transition and Emergent Fractionalized Supersymmetry in the Non-Abelian Kitaev Spin Liquid''}
\end{center}

\renewcommand{\theequation}{S\arabic{equation}}
\renewcommand{\thefigure}{S\arabic{figure}}
\renewcommand{\thetable}{S\arabic{table}}
\renewcommand\figurename{Supplemental Figure}
\renewcommand\tablename{Supplemental Table}

\section{Vectorization conventions and fractionalized Hamiltonian}

\subsection{Vectorization convention}

At each site $i$, we introduce the maximally entangled state
\begin{equation}
 \lvert\Omega_i\rangle\!\rangle
 =\tfrac{1}{\sqrt{2}}\left(\lvert\uparrow\rangle_L\lvert\uparrow\rangle_R
 +\lvert\downarrow\rangle_L\lvert\downarrow\rangle_R \right).
 \label{eqS:omega-site}
\end{equation}
The normalized many-body reference state is
$\lvert\Omega\rangle\!\rangle=\prod_i\lvert\Omega_i\rangle\!\rangle$.
We vectorize an operator $O$ according to
\begin{equation}
 \lvert O\rangle\!\rangle=(O\otimes\mathbbm 1)\lvert\Omega\rangle\!\rangle,
 \qquad
 O_L\lvert\Omega\rangle\!\rangle=O_R^T\lvert\Omega\rangle\!\rangle,
 \label{eqS:vectorization}
\end{equation}
where the transpose is taken in the product $\sigma^z$ basis.
In this convention,
\begin{equation}
 \sigma_L^x\lvert\Omega\rangle\!\rangle=\sigma_R^x\lvert\Omega\rangle\!\rangle,
 \quad
 \sigma_L^y\lvert\Omega\rangle\!\rangle=-\sigma_R^y\lvert\Omega\rangle\!\rangle,
 \quad
 \sigma_L^z\lvert\Omega\rangle\!\rangle=\sigma_R^z\lvert\Omega\rangle\!\rangle.
 \label{eqS:pauli-transpose}
\end{equation}
The bond-dephasing channel acts on the density matrix as
\begin{equation}
\mathcal E_{ij}^\alpha(\rho)
=(1-p_\alpha)\rho
+p_\alpha B_{ij}^\alpha\rho B_{ij}^\alpha,
\qquad
B_{ij}^\alpha=\sigma_i^\alpha\sigma_j^\alpha.
\label{eqS:kraus}
\end{equation}
Under vectorization,
$\lvert\mathcal E_{ij}^\alpha(\rho)\rangle\!\rangle
=\widehat{\mathcal E}_{ij}^\alpha
\lvert\rho\rangle\!\rangle$,
where the channel is represented in the doubled Hilbert space by
\begin{equation}
 \widehat{\mathcal E}_{ij}^\alpha
 =(1-p_\alpha)\mathbbm 1+p_\alpha D_{ij}^\alpha,
 \qquad D_{ij}^\alpha=B_{ij,L}^\alpha B_{ij,R}^\alpha .
 \label{eqS:single-channel}
\end{equation}
Here we have used $(B_{ij}^\alpha)^T=B_{ij}^\alpha$
in the chosen spin basis.
Using $(D_{ij}^\alpha)^2=\mathbbm{1}$, the single-bond channel can be written exactly as
\begin{equation}
 \widehat{\mathcal E}_{ij}^\alpha=e^{-V_\alpha}
 e^{V_\alpha D_{ij}^\alpha},
 \qquad
 e^{-2V_\alpha}=1-2p_\alpha.
 \label{eqS:channel-exponential}
\end{equation}
All doubled bond operators commute: any anticommutation sign in one copy is canceled by the same sign in the other.
The full channel therefore has the exact representation
\begin{equation}
 \widehat{\mathcal E}=\prod_{\alpha,\langle ij\rangle_\alpha}
 \widehat{\mathcal E}_{ij}^\alpha
 \propto\exp\!\left(\sum_{\alpha,\langle ij\rangle_\alpha}
 V_\alpha D_{ij}^\alpha\right)
 \label{eqS:full-channel}
\end{equation}
The proportionality omits only an overall scalar;
no Trotter approximation is involved.

The channel also preserves every plaquette flux.
To establish this directly in the physical spin Hilbert space, label the vertices of a hexagon $h$ cyclically by $n=1,\ldots,6$, and let $\beta_n$ denote the type of the bond leaving the plaquette at vertex $n$.
The flux operator can be defined as
\begin{equation}
 W_h=\prod_{n=1}^{6}\sigma_n^{\beta_n}.
 \label{eqS:spin-flux}
\end{equation}
For a conventional cyclic labeling, $W_h=\sigma_1^x\sigma_2^y\sigma_3^z\sigma_4^x\sigma_5^y\sigma_6^z$.
A bond not on $\partial h$ either has disjoint support from $W_h$ or is the outgoing bond at its only shared vertex, where the Pauli components coincide.
For a bond on $\partial h$, the Pauli components differ at both endpoints, so the two anticommutation signs cancel.
Thus, for every bond and plaquette,
\begin{equation}
 [B_{ij}^\alpha,W_h]=0.
 \label{eqS:spin-flux-commutation}
\end{equation}
The channel also preserves every flux sector.
In the doubled Hilbert space, the fluxes of the two copies
are conserved independently:
\begin{equation}
    [D_{ij}^\alpha, W_{h,L}] = [D_{ij}^\alpha, W_{h,R}] = 0.
\end{equation}

\subsection{Majorana representation of the doubled system}

We now formulate the left and right copies appearing after vectorization, together with the flux-preserving dephasing channel, in the Majorana representation.
Following Ref.~\cite{kitaev2006anyons}, we introduce four Majoranas
at each site in each copy,
\begin{equation}
 \sigma_{i,s}^\alpha=\ii b_{i,s}^\alpha c_{i,s},
 \qquad
 G_{i,s}=b_{i,s}^xb_{i,s}^yb_{i,s}^zc_{i,s}=+1,
 \qquad s=L,R.
 \label{eqS:majorana-spin}
\end{equation}
with the physical subspace defined by $G_{i,s}=+1$.
With a slight abuse of notation, we also use
$\lvert\Omega\rangle\!\rangle$ for an unprojected reference state
in the enlarged Majorana Hilbert space, chosen to satisfy
\begin{align}
 b_L^x\lvert\Omega\rangle\!\rangle&=\ii b_R^x\lvert\Omega\rangle\!\rangle,&
 b_L^y\lvert\Omega\rangle\!\rangle&=-\ii b_R^y\lvert\Omega\rangle\!\rangle,\nonumber\\
 b_L^z\lvert\Omega\rangle\!\rangle&=\ii b_R^z\lvert\Omega\rangle\!\rangle,&
 c_L\lvert\Omega\rangle\!\rangle&=-\ii c_R\lvert\Omega\rangle\!\rangle.
 \label{eqS:majorana-copy-map}
\end{align}
These phase conventions reproduce the spin-space vectorization identities in Eq.~\eqref{eqS:pauli-transpose} through $\sigma^\alpha=\ii b^\alpha c$.

We introduce the $\mathbb Z_2$ link operator, defined on bonds by
\begin{equation}
 u_{ij,s}^\alpha=-u_{ji,s}^\alpha=\ii\zeta_s b_{i,s}^\alpha b_{j,s}^\alpha,
\qquad\zeta_L = -\zeta_R=1.
 \label{eqS:links}
\end{equation}
The factor $\zeta_s$ compensates for the relative minus sign between the bond bilinears under Eq.~\eqref{eqS:majorana-copy-map}, so that corresponding links obey $u_{ij,L}^\alpha\lvert\Omega\rangle\!\rangle
=u_{ij,R}^\alpha\lvert\Omega\rangle\!\rangle$.
The exchange operator then becomes
\begin{equation}
 B_{ij,s}^\alpha=\sigma_{i,s}^\alpha\sigma_{j,s}^\alpha
 =-\ii\zeta_s u_{ij,s}^\alpha c_{i,s}c_{j,s}.
 \label{eqS:bond-majorana}
\end{equation}
Using $\zeta_L\zeta_R=-1$, the doubled bond operator becomes
\begin{equation}
 D_{ij}^\alpha=u_{ij,L}^\alpha u_{ij,R}^\alpha
 c_{i,L}c_{j,L}c_{i,R}c_{j,R}.
 \label{eqS:quartic-channel}
\end{equation}
For an oriented two-bond path $i\xrightarrow{\alpha}j\xrightarrow{\beta}k$, let $\gamma$ denote the remaining spin component.
Using the physical constraint $G_{j,s}=1$, the three-spin
interaction can be written as
\begin{equation}\sigma_{i,s}^\alpha\sigma_{j,s}^\gamma\sigma_{k,s}^\beta
 =\ii\epsilon_{\alpha\beta\gamma}
 u_{ij,s}^\alpha u_{jk,s}^\beta c_{i,s}c_{k,s},
 \label{eqS:three-spin-majorana}
\end{equation}
where $\epsilon_{\alpha\beta\gamma}$ is the Levi-Civita symbol with $\epsilon_{xyz}=+1$.

All link operators commute with one another, with each term in the Hamiltonian, and with the dephasing channel.
We may therefore work in a fixed configuration $u_{ij,s}^\alpha=\pm1$, subject to the zero-flux constraint
\begin{equation}
 W_{h,s}
 =\prod_{\langle ij\rangle\in\partial h}
 u_{ij,s}^{\alpha_{ij}}=1
\end{equation}
on every hexagon, with each link in the product taken in the
fixed $A$-to-$B$ orientation.
Throughout this work, we choose $u_{ij,s}^\alpha=1$ on $A$-to-$B$ bonds in the bulk of both copies.
On a finite torus, the matter states must also satisfy the physical-parity constraint~\cite{pedrocchi2011physical}.
For the odd-$L$ clusters used here, we impose a $\pi$ boundary twist along one primitive direction.
The exchange couplings change sign on bonds crossing that
boundary, and each three-spin coupling acquires the product
of the bond signs along its two-bond path.
Taking $u_{ij,s}^\alpha=-1$ on these boundary-crossing bonds
cancels the coupling signs, so that the matter Hamiltonians
remain periodic.
This choice preserves $W_h=1$ and makes both the topological
and reference vacua compatible with physical projection.

With the gauge and boundary conventions above, Eqs.~\eqref{eqS:bond-majorana} and~\eqref{eqS:three-spin-majorana} yield the periodic matter Hamiltonians
\begin{equation}
 H_s=\zeta_s \left( \ii\sum_{\alpha,\langle ij\rangle_\alpha}J_\alpha c_{i,s}c_{j,s}
 +\ii\kappa\sum_{\langle\!\langle ik\rangle\!\rangle}
 \nu_{ik}c_{i,s}c_{k,s} \right),
 \label{eqS:uniform-majorana-hamiltonian}
\end{equation}
where $\nu_{ik}=\epsilon_{\alpha\beta\gamma}=-\nu_{ki}$ for the oriented path defined above.
The right-copy spin Hamiltonian is the transpose of the left-copy Hamiltonian: the exchange terms are unchanged, whereas each three-spin term changes sign because it contains exactly one $\sigma^y$.
Together with the copy-dependent link convention, this accounts for the common prefactor $\zeta_s$. 
The two copies consequently have opposite matter-band Chern numbers.

Finally, define the inter-copy bilinear $\Phi_i=\ii c_{i,L}c_{i,R}$.
For isotropic dephasing, Eqs.~\eqref{eqS:quartic-channel}
and~\eqref{eqS:full-channel} give
\begin{equation}
 D_{ij}^\alpha=\Phi_i\Phi_j,
 \qquad
 \widehat{\mathcal E}(p)\propto
 \exp\!\left[V\sum_{\langle ij\rangle}\Phi_i\Phi_j\right].
 \label{eqS:ferromagnetic-channel}
\end{equation}

\subsection{Physical completion of the parton correlators}
The Majoranas $c_{i,s}$ and the inter-copy bilinear $\Phi_i$ carry gauge charge.
To relate their fixed-gauge correlators to physical spin observables, we connect the endpoints by gauge Wilson lines.
For an oriented path $\gamma$ from site $i$ to site $j$, define
\begin{equation}
 U_{\gamma,s}=\prod_{\langle lm\rangle\in\gamma}u_{lm,s}.
 \label{eqS:wilson-line}
\end{equation}
Under a local gauge transformation, the signs associated with internal vertices cancel, leaving only those at the endpoints.
These endpoint signs are canceled by the matter Majoranas, giving the gauge-invariant fermion string
\begin{equation}
 \mathcal W_{f,s}(i,j;\gamma)
 =\ii c_{i,s}U_{\gamma,s}c_{j,s}.
 \label{eqS:fermion-string}
\end{equation}
Since $\Phi_i$ carries gauge charge in both copies, its two-point function requires a Wilson line in each copy:
\begin{equation}
 \mathcal W_b(i,j;\gamma)=
 \Phi_i U_{\gamma,L}U_{\gamma,R}\Phi_j.
 \label{eqS:boson-string}
\end{equation}
These Wilson strings have a direct spin representation.
For a path $\gamma$, define
the spin string
\begin{equation}
 W_{\gamma,s}
 \equiv\prod_{(lm)_\alpha\in\gamma}B_{lm,s}^\alpha
 =s_{\gamma,s}
 \mathcal W_{f,s}(i,j;\gamma).
 \label{eqS:spin-string}
\end{equation}
The product is ordered along $\gamma$ from $i$ to $j$.
For a path containing $\ell_\gamma$ bonds, Eq.~\eqref{eqS:bond-majorana} fixes the phase to $s_{\gamma,s}=-\ii(-\ii\zeta_s)^{\ell_\gamma}$.
The phase product $s_{\gamma,L}s_{\gamma,R}=-1$ cancels
the fermionic reordering sign in $\mathcal W_{f,L}\mathcal W_{f,R}=-\mathcal W_b$, giving $W_{\gamma,L}W_{\gamma,R}=\mathcal W_b$.

The spin strings defined above also identify the corresponding
density-matrix observables.
Since $W_\gamma$ is real and unitary in the basis used for
vectorization, $W_\gamma^T=W_\gamma^\dagger$.
Its insertion a normalized Choi state
therefore gives the R\'enyi-2 string correlator
\begin{equation}
 C_2(W_\gamma;\rho)
 =
 \frac{\langle\!\langle\rho\rvert
 W_{\gamma,L}W_{\gamma,R}
 \lvert\rho\rangle\!\rangle}
 {\langle\!\langle\rho\vert\rho\rangle\!\rangle}
 =
 \frac{\Tr(\rho W_\gamma\rho W_\gamma^\dagger)}
 {\Tr(\rho^2)}.
 \label{eqS:renyi-string}
\end{equation}
For the dephased state, we evaluate
$C_2(W_\gamma;\rho_p)$ with $\lvert\rho_p\rangle\!\rangle
=\widehat{\mathcal E}(p)\lvert\rho_0\rangle\!\rangle$.
Since $\widehat{\mathcal E}(p)$ is Hermitian and the doubled Wilson string commutes with the channel, these factors can be combined.
Using $D_{ij}^2=1$ and the mutual commutativity of the doubled
bond operators, we obtain
\begin{equation}
 \widehat{\mathcal E}(p)^2=\widehat{\mathcal E}(q),
 \qquad q=2p(1-p).
 \label{eqS:two-layer-conversion}
\end{equation}
The correlator can therefore be expressed in terms of the
initial projector as
\begin{equation}
 C_2(W_\gamma;\rho_p)=
 \frac{\langle\!\langle\rho_0\rvert\widehat{\mathcal E}(p)
 W_{\gamma,L}W_{\gamma,R}\widehat{\mathcal E}(p)
 \lvert\rho_0\rangle\!\rangle}
 {\langle\!\langle\rho_0\rvert\widehat{\mathcal E}(p)^2
 \lvert\rho_0\rangle\!\rangle}=
 \frac{\langle\!\langle\rho_0\rvert
 W_{\gamma,L}W_{\gamma,R}\widehat{\mathcal E}(q)
 \lvert\rho_0\rangle\!\rangle}
 {\langle\!\langle\rho_0\rvert
 \widehat{\mathcal E}(q)
 \lvert\rho_0\rangle\!\rangle}.
 \label{eqS:bulk-renyi-two}
\end{equation}
Here $p$ specifies the physical output state $\rho_p$, while
$q$ parametrizes the combined channel in this representation.
Long-range order in Eq.~\eqref{eqS:bulk-renyi-two} demonstrates $\psi\bar\psi$ condensation in the normalized Choi state.
The imTO identification does not use this correlator as a phase definition: it follows by unfolding the condensed-anyon structure at maximal dephasing, with the finite-$p$ assignment assuming that no additional transition intervenes within the ordered branch~\cite{sang2023mixed,sohal2025noisy}.

For the topological interface, the distinct topological and reference boundaries instead give the doubled-string amplitude
\begin{equation}
 C_b^{\rm str}(i,j;\gamma)=
 \frac{\Tr\!\left[\rho_{\rm ref}W_\gamma
 \mathcal E_p(\rho_{\rm top})W_\gamma^\dagger\right]}
 {\Tr\!\left[\rho_{\rm ref}\mathcal E_p(\rho_{\rm top})\right]}=
 \frac{\langle\!\langle\rho_{\rm ref}\rvert
 W_{\gamma,L}W_{\gamma,R}\widehat{\mathcal E}(p)
 \lvert\rho_{\rm top}\rangle\!\rangle}
 {\langle\!\langle\rho_{\rm ref}\rvert
 \widehat{\mathcal E}(p)
 \lvert\rho_{\rm top}\rangle\!\rangle}.
 \label{eqS:boson-strange-amplitude}
\end{equation}
This is a type-II strange correlator because the physical string acts on both copy indices.
The fermion string enters the corresponding normalized transition amplitude,
\begin{equation}
 C_f^{\rm str}(i,j;\gamma)=
 \frac{\langle\!\langle\rho_{\rm ref}\rvert
 W_{\gamma,L}\widehat{\mathcal E}(p)
 \lvert\rho_{\rm top}\rangle\!\rangle}
 {\langle\!\langle\rho_{\rm ref}\rvert
 \widehat{\mathcal E}(p)
 \lvert\rho_{\rm top}\rangle\!\rangle}.
 \label{eqS:fermion-transition-amplitude}
\end{equation}
This is the corresponding type-I strange correlator with a string insertion in one copy only~\cite{lee2025symmetry}.
At the supersymmetric point,
$C_b^{\rm str}(r)\sim r^{-2/5}$ and $C_f^{\rm str}(r)\sim r^{-7/5}$.
The equality is between their anomalous parts, not their full powers, because the fermion carries the canonical factor $r^{-1}$.

\section{Sign-problem-free Quantum Monte Carlo and simulation details}

\subsection{Auxiliary-field decomposition and nonnegative weights}
For isotropic dephasing, the single-bond channel admits the exact discrete Hubbard--Stratonovich representation
\begin{equation}
\widehat{\mathcal E}_{ij}(p)=e^{-V}e^{VD_{ij}^\alpha}=e^{-\frac{V}{2}
(\ii c_{i,L}c_{j,L}+\ii c_{i,R}c_{j,R})^2}=\frac{1}{2}\sum_{s_{ij}=\pm1}
e^{\theta s_{ij}
(c_{i,L}c_{j,L}+c_{i,R}c_{j,R})},
\quad
\cos(2\theta)=e^{-2V}=1-2p.
\label{eqS:hs-identity}
\end{equation}
Here $s_{ij}=\pm1$ is an auxiliary Ising field on each bond.
Since the interactions $D_{ij}^\alpha$ commute on different bonds, the full channel factorizes exactly into single-bond channels.
Within each bond-type layer $\alpha=x,y,z$, the bonds are disjoint, so their Majorana bilinears also commute and can be grouped into a single Gaussian exponential.
Applying Eq.~\eqref{eqS:hs-identity} to every bond gives
\begin{equation}
\widehat{\mathcal E}(p)
=\frac{1}{2^{N_b}}\sum_{\mathcal C}
\prod_{\alpha=x,y,z}^{\rightarrow}
\exp\!\left[
\theta\sum_{\langle ij\rangle_\alpha}
s_{ij}\left(c_{i,L}c_{j,L}+c_{i,R}c_{j,R}\right)\right],
\label{eqS:full-hs}
\end{equation}
where $N_b$ is the total number of bonds,
$\mathcal C=\{s_{ij}\}$ denotes an auxiliary-field configuration,
and the arrow specifies a fixed ordering of the three bond layers.
For a fixed $\mathcal C$, Gaussian factors in different layers
generally do not commute, so this ordering is retained explicitly.
The auxiliary-field sum nevertheless reproduces the channel exactly,
with no Trotter approximation.

Let $\mathcal G_s(\mathcal C)$ denote the ordered product of Gaussian factors acting on copy $s$ in Eq.~\eqref{eqS:full-hs}.
We consider boundary states $\kket{\rho_a}$, $\kket{\rho_b}$, with complex-conjugate partners in the two copies,
$|\psi_{a,R}\rangle=|\psi_{a,L}^*\rangle$ and similarly for $b$.
The channel overlap between these doubled boundary states is
$Z_{ab}=\sum_{\mathcal C}w_{ab}(\mathcal C)$, where
\begin{equation}
  w_{ab}(\mathcal C)
  =\frac{1}{2^{N_b}}
  \prod_{s=L,R}
  \langle\psi_{a,s}|
  \mathcal G_s(\mathcal C)
  |\psi_{b,s}\rangle.
  \label{eqS:configuration-weight}
\end{equation}
For a fixed auxiliary-field configuration, the Hubbard--Stratonovich factors preserve the conjugate relation between the two copies.
Consequently, $\langle\psi_{a,R}|\mathcal G_R(\mathcal C)|\psi_{b,R}\rangle=\left[\langle\psi_{a,L}|\mathcal G_L(\mathcal C)|\psi_{b,L}\rangle\right]^*$,
and hence
\begin{equation}
  w_{ab}(\mathcal C)
  =\frac{1}{2^{N_b}}
  \left|
  \langle\psi_{a,L}|
  \mathcal G_L(\mathcal C)
  |\psi_{b,L}\rangle
  \right|^2
  \ge0.
  \label{eqS:nonnegative-weight}
\end{equation}
This conjugate pairing eliminates the fermion sign problem
configuration by configuration~\cite{li2015solving,li2016majorana,wei2016majorana}.
Configurations are sampled with probability
$w_{ab}(\mathcal C)/Z_{ab}$.

\subsection{Gaussian overlaps and correlation estimators}
To evaluate the single-copy overlap
$A_{ab}(\mathcal C)
=\langle\psi_{a,L}|\mathcal G_L(\mathcal C)|\psi_{b,L}\rangle$,
we write
\begin{equation}
  \mathcal G_L(\mathcal C)
  =\prod_{\alpha=x,y,z}^{\rightarrow}
  \exp\!\left[
  \frac14\boldsymbol c_L^T
  K_\alpha(\mathcal C)\boldsymbol c_L
  \right].
  \label{eqS:single-copy-gaussian}
\end{equation}
Here $K_\alpha$ is an antisymmetric matrix with
$(K_\alpha)_{ij}=-(K_\alpha)_{ji}=2\theta s_{ij}$
on each $\alpha$ bond directed from $A$ to $B$,
and all other entries zero.
Following Ref.~\cite{li2015solving}, we introduce an identical auxiliary Majorana copy $\widetilde c_{i,L}$, distinct from the
vectorization copy $R$, and define
$f_i=(c_{i,L}+\ii\widetilde c_{i,L})/2$.
The two identical Gaussian generators then combine into a number-conserving form,
\begin{equation}
  \frac14\boldsymbol c_L^T K_\alpha\boldsymbol c_L
  +\frac14\widetilde{\boldsymbol c}^{\,T}
  K_\alpha\widetilde{\boldsymbol c}
  =\boldsymbol f^\dagger K_\alpha\boldsymbol f.
  \label{eqS:auxiliary-majorana-pairing}
\end{equation}
Likewise, pairing each boundary state with its identical auxiliary
copy gives a Slater determinant
$|F_a\rangle=|\psi_{a,L}\rangle\otimes|\widetilde\psi_{a,L}\rangle$.
Let $P_a$ denote its $N\times(N/2)$ occupied-orbital matrix,
and similarly for $P_b$.
For ground-state boundaries, these orbitals are obtained by
diagonalizing the corresponding duplicated boundary Hamiltonian.
The squared overlap becomes
\begin{equation}
    A_{ab}(\mathcal C)^2=
    \langle F_a|
    \prod_{\alpha=x,y,z}^{\rightarrow}
    e^{\boldsymbol f^\dagger K_\alpha(\mathcal C)\boldsymbol f}
    |F_b\rangle=\det M_{ab}(\mathcal C),\quad M_{ab}(\mathcal C)=P_a^\dagger\left[\prod_{\alpha=x,y,z}^{\rightarrow}e^{K_\alpha(\mathcal C)}\right]P_b.
  \label{eqS:gaussian-overlap-determinant}
\end{equation}
Using $|A_{ab}|^2=|A_{ab}^2|$, the nonnegative sampling weight
therefore takes the determinant form
\begin{equation}
  w_{ab}(\mathcal C)
  =\frac{1}{2^{N_b}}
  \left|\det M_{ab}(\mathcal C)\right|.
  \label{eqS:determinant-weight}
\end{equation}

The determinant representation allows standard determinant quantum Monte Carlo
techniques to be used for matrix updates and Green-function evaluation.
For each auxiliary-field configuration, we define the normalized
transition Green function
\begin{equation}
  g_{ij,s}(\mathcal C)
  =
  \frac{
  \langle\psi_{a,s}|
  c_{i,s}c_{j,s}\mathcal G_s(\mathcal C)
  |\psi_{b,s}\rangle
  }{
  \langle\psi_{a,s}|
  \mathcal G_s(\mathcal C)
  |\psi_{b,s}\rangle
  }.
  \label{eqS:transition-green}
\end{equation}
Using the Slater-determinant representation, the transition
Green function is
\begin{equation}
  g_{ij,L}(\mathcal C)
  =
  2\left[
  \left(\prod_{\alpha=x,y,z}^{\rightarrow}
  e^{K_\alpha(\mathcal C)}\right)
  P_b M_{ab}(\mathcal C)^{-1}P_a^\dagger
  \right]_{ji},
  \qquad
  g_{ij,R}(\mathcal C)=g_{ij,L}(\mathcal C)^*.
  \label{eqS:transition-green-matrix}
\end{equation}
The factor of two follows from
$\langle f_i^\dagger f_j\rangle_{\mathcal C}
=g_{ij,L}(\mathcal C)/2$
in the identical auxiliary-copy construction.

Since the two copies decouple at fixed $\mathcal C$,
the bosonic and fermionic correlators are obtained as
\begin{equation}
  \begin{aligned}
    \langle\Phi_i\Phi_j\rangle_{ab}
    &=
    \left\langle
    g_{ij,L}(\mathcal C)g_{ij,R}(\mathcal C)
    \right\rangle_{\mathrm{MC}}
    =
    \left\langle
    |g_{ij,L}(\mathcal C)|^2
    \right\rangle_{\mathrm{MC}},\\
    \langle c_{i,L}c_{j,L}\rangle_{ab}
    &=
    \left\langle g_{ij,L}(\mathcal C)\right\rangle_{\mathrm{MC}}.
  \end{aligned}
  \label{eqS:correlation-estimators}
\end{equation}
Here $\langle\cdots\rangle_{\mathrm{MC}}$ denotes averaging with
probability $w_{ab}(\mathcal C)/Z_{ab}$.
Higher-order correlation functions are evaluated using Wick's
theorem at fixed $\mathcal C$ before Monte Carlo averaging.
For the normalized output Choi state, the same bosonic estimator
applies with identical initial boundary states and effective
channel probability $q=2p(1-p)$ as shown in Eq.~\eqref{eqS:bulk-renyi-two}.

\subsection{Observables and finite size scaling}
In the simulation, we use periodic honeycomb clusters with $L\times L$ unit cells, $N=2L^2$ sites.
The primitive and reciprocal vectors are
\begin{equation}
  \mathbf a_{1,2}
  =\left(\frac12,\pm\frac{\sqrt3}{2}\right),
  \qquad
  \mathbf b_{1,2}
  =2\pi\left(1,\pm\frac1{\sqrt3}\right),
  \label{eqS:lattice-vectors}
\end{equation}
with $\mathbf a_\mu\cdot\mathbf b_\nu=2\pi\delta_{\mu\nu}$.
We label sites by a unit-cell coordinate
$\mathbf R=n_1\mathbf a_1+n_2\mathbf a_2$
and a sublattice index $\mu=A,B$.
Allowed momenta are
$\mathbf q=(m_1\mathbf b_1+m_2\mathbf b_2)/L$,
and we choose $\mathbf q_{\min}=\mathbf b_1/L$.

Using the bosonic correlators defined above, the same-sublattice
structure factor is
\begin{equation}
  S_{\Phi,\mu\mu}(\mathbf q)
  =
  \frac1{L^2}\sum_{\mathbf R,\mathbf R'}
  e^{\ii\mathbf q\cdot(\mathbf R-\mathbf R')}
  \langle\Phi_{\mathbf R,\mu}\Phi_{\mathbf R',\mu}\rangle.
  \label{eqS:structure-factor}
\end{equation}
The squared order parameter and dimensionless correlation ratio are
\begin{equation}
  m_{\Phi,\mu\mu}^2
  =\frac{S_{\Phi,\mu\mu}(\mathbf0)}{L^2},
  \qquad
  R_{\Phi,\mu\mu}
  =1-\frac{S_{\Phi,\mu\mu}(\mathbf q_{\min})}
           {S_{\Phi,\mu\mu}(\mathbf0)}.
  \label{eqS:bosonic-observables}
\end{equation}
These definitions apply to both the normalized-output and
temporal-interface calculations, using their respective correlators.

For the temporal interface, we additionally measure the
window-averaged fermionic correlation
\begin{equation}
  \overline{\mathcal{C}}_{f,\mu\mu}^{\mathrm{win}}
  =
  \frac1{|\mathcal W_L|}
  \sum_{\mathbf d\in\mathcal W_L}
  \left|
  \frac1{L^2}\sum_{\mathbf R}
  \langle
  c_{\mathbf R,\mu,L}c_{\mathbf R+\mathbf d,\mu,L}
  \rangle_{\mathrm{str}}
  \right|.
  \label{eqS:fermion-window}
\end{equation}
The window $\mathcal W_L$ contains displacements
$\mathbf d=d_1\mathbf a_1+d_2\mathbf a_2$
with integer $L/3\le d_1,d_2\le2L/3$.
The absolute value is taken after averaging the correlator,
and the window selects separations proportional to $L$.

All simulations are performed at $J_x=J_y=1$ and $\kappa=0.5$, with isotropic dephasing.
We discard the first 500 sweeps of each independent Markov chain for equilibration.
Each sweep updates all bonds in the fixed $x,y,z$ layer order, followed by measurements.
Green functions are restabilized every five sweeps.
We analyze the critical behavior using the finite-size scaling forms
\begin{equation}
\begin{split}
R_\Phi
&=F_R(x)+L^{-\omega}G_R(x),\\
m_\Phi^2
&=L^{-\eta_b}\left[F_b(x)+L^{-\omega}G_b(x)\right],\\
\overline{\mathcal C}_f^{\rm win}
&=L^{-1-\eta_f}\left[F_f(x)+L^{-\omega}G_f(x)\right].
\label{eqS:scaling-form}
\end{split}
\end{equation}
Here $x=(p-p_c)L^{1/\nu}$, and the $AA$ sublattice subscripts are omitted.
Other sublattice channels exhibit consistent critical behavior.
The scaling functions $F_a$ and correction functions $G_a$ are approximated by polynomials of degree $m$ and $n$, respectively. The correction exponent $\omega$ is specified separately
for each observable below.

For the Choi-state calculation, we take $J_z=1$ for both boundary states.
The bosonic string operators commute with the channel, allowing the two channel factors in the correlator to combine as $\widehat{\mathcal E}(p)^2=\widehat{\mathcal E}(q)$, where $q=2p(1-p)$ and $p$ is the physical dephasing probability.
We fit the data for $21\leq L\leq51$ and $|p-0.167|\leq0.009$, using $m=7$ and $n=2$.
The correlation-ratio correction exponent is poorly constrained, so we fix $\omega=2$ to stabilize the fit; the bosonic correction exponent is fitted independently.
The correlation-ratio fit gives $p_c^{(2)}=0.16687(8)$ and $\nu=1.011(16)$, with $\chi^2/{\rm dof}=0.655$.
Fixing these two parameters in the subsequent fit of $m_\Phi^2$ yields $\eta=0.246(11)$,
with $\chi^2/{\rm dof}=1.031$. These estimates are consistent with the two-dimensional Ising exponents $\nu=1$ and $\eta=1/4$.

For the topological-interface calculation, we choose $J_z=3$ and
$J_z=1$ for the bra and ket boundary states, respectively.
The data in Fig.~3(a)--(c) are obtained from $(3.20\text{--}3.36)\times10^5$ Monte Carlo sweeps per point.
We supplement these data with independent simulations at $p=0.23,0.2451513,0.25$ and $L=21,27,\ldots,57$, using 32--384 Markov chains and a total of $(1.68\text{--}1.73)\times10^6$ sweeps per point.
All sweep counts refer to sampling after equilibration and are summed over chains.
We fit the combined data for $21\leq L\leq57$ and $|p-0.2452|\leq0.012$, using $m=6$ and $n=2$.
As in the Choi-state analysis, we fix $\omega=2$ for the correlation ratio and allow the bosonic and fermionic correction exponents to vary independently.
The resulting estimates are $p_c=0.24526(6)$, $\nu=1.27(3)$, $\eta_b=0.405(6)$, and $\eta_f=0.395(21)$.
The corresponding $\chi^2/{\rm dof}$ values are $57.56/50$, $55.57/50$, and $62.77/50$ for the correlation-ratio, bosonic, and fermionic fits, respectively.

To compare the spatial anisotropies of the two sectors, we evaluate the translation-averaged $AA$ correlators $C_a(\mathbf d)$, with $a=b,f$, at $\mathbf d_\pm=\frac L3(\mathbf a_1\pm\mathbf a_2)$.
For the lattice convention used here,
$\mathbf d_+=(L/3,0)$ and
$\mathbf d_-=(0,\sqrt3L/3)$.
Motivated by the anisotropic power-law form $|C_a(x,y)|\propto[x^2+(y/\lambda_a)^2]^{-\alpha_a/2}$, we define
\begin{equation}
  \lambda_a^{\mathrm{eff}}(L)
  =
  \sqrt3\left|
  \frac{C_a(\mathbf d_-)}{C_a(\mathbf d_+)}
  \right|^{1/\alpha_a},
  \qquad
  \alpha_b=\frac25,\quad \alpha_f=\frac75.
  \label{eqS:anisotropy-estimator}
\end{equation}
The powers are fixed to the supersymmetric predictions.
Since the separations remain fixed fractions of $L$, $\lambda_a^{\rm eff}$ provides a finite-size measure of correlation anisotropy.
We use the high-statistics data at $p=0.23,p_*,0.25$, where $p_*=0.2451513$ denotes the near-critical simulation point.
The relative anisotropy is quantified by
\begin{equation}
\Delta_\lambda(L)
=\log\frac{\lambda_f^{\rm eff}(L)}
{\lambda_b^{\rm eff}(L)},
\qquad
\Delta_\lambda(L)
=\Delta_\lambda^{(\infty)}+\frac{A_\lambda}{L}.
\label{eqS:anisotropy-extrapolation}
\end{equation}
At each probability, we fit the data for $L=33,39,\ldots,57$ with a free intercept.
Error bars are obtained from a paired delete-one-chain jackknife applied to the full estimator.
For the near-critical dataset, the extrapolation gives $\Delta_\lambda^{(\infty)}=0.006(19)$,
consistent with vanishing boson--fermion anisotropy mismatch.

\section{From doubled-Ising order to intrinsic mixed-state topological order}

\subsection{Condensation in the normalized Choi state}

The chiral non-Abelian phase of one Kitaev copy has Ising anyons
$\{1,\sigma,\psi\}$, quantum dimensions
$d_1=d_\psi=1$, $d_\sigma=\sqrt2$, and fusion rules
\begin{equation}
 \psi\times\psi=1,
 \qquad \psi\times\sigma=\sigma,
 \qquad \sigma\times\sigma=1+\psi.
 \label{eqS:ising-fusion}
\end{equation}
Their topological spins are
$\theta_1=1$, $\theta_\psi=-1$, and $\theta_\sigma=e^{\ii\pi/8}$~\cite{kitaev2006anyons,nayak2008nonabelian}.
Vectorizing a pure density matrix produces the nonchiral product
$\Ising\boxtimes\overline{\Ising}$, whose nine sectors are denoted $a\bar b$.

The channel interaction creates the inter-copy pair
$b_c=\psi\bar\psi$.
It is an Abelian boson because
$d_{b_c}=1$ and
$\theta_{b_c}=\theta_\psi\theta_{\bar\psi}=1$.
An anyon remains deconfined after condensation only if its monodromy with $b_c$ is trivial.
The five sectors satisfying this condition are
\begin{equation}
 1\bar1,\quad \psi\bar\psi,\quad
 \psi\bar1,\quad1\bar\psi,\quad\sigma\bar\sigma.
 \label{eqS:surviving-sectors}
\end{equation}
The other four sectors with exactly one $\sigma$ have nontrivial monodromy and are confined.

Fusion with the condensate produces the identifications
\begin{equation}
 1\bar1\sim\psi\bar\psi\equiv1,
 \qquad
 \psi\bar1\sim1\bar\psi\equiv\epsilon.
 \label{eqS:identification}
\end{equation}
The sector $\sigma\bar\sigma$ is a fixed point under fusion with $b_c$ and has quantum dimension two, so it must split into two Abelian sectors,
\begin{equation}
 \sigma\bar\sigma\longrightarrow e+m.
 \label{eqS:splitting}
\end{equation}
This splitting is also required by its fusion square:
\begin{align}
    (\sigma\bar\sigma)^2=(1+\psi)(\bar1+\bar\psi)=1\bar1+\psi\bar\psi+\psi\bar1+1\bar\psi
    \longrightarrow 2(1+\epsilon),
\end{align}
which equals $(e+m)^2$ for toric-code fusion.
The descendants obey $e^2=m^2=\epsilon^2=1$ and $e\times m=\epsilon$, with $\theta_e=\theta_m=1$ and $\theta_\epsilon=-1$.
The total quantum dimension changes from $\mathcal D_{\Ising\boxtimes\overline{\Ising}}=4$ to $\mathcal D_{\rm TC}=2$, as expected after condensation of an algebra of dimension two~\cite{bais2009condensate,kong2014anyon,burnell2018anyon,sohal2025noisy}.

\subsection{Unfolding to the density matrix}

The doubled-state result cannot be interpreted literally as a physical bilayer.
Under the inverse Choi map,
\begin{equation}
 a\bar b \lvert \rho \rangle\!\rangle \quad\longleftrightarrow\quad W_a\rho W_b^\dagger,
 \label{eqS:unfold-sector}
\end{equation}
where $W_a$ and $W_b$ are string operators carrying the corresponding topological charges.
At the maximal-dephasing fixed point, doubled condensation $\psi \bar \psi$ makes the mixed state invariant under inserting a closed $\psi$ string on the ket and bra simultaneously.
This operation gauges out $\psi$ in operator space rather than erasing the distinction between a one-sided $\psi$ and the vacuum~\cite{sohal2025noisy}.

After unfolding, the deconfined one-sided content is generated by $\{1,\psi\}$.
Because $\psi$ braids trivially with every surviving sector while retaining fermionic spin, it lies in the transparent center of the braided category.
The category is therefore nonmodular, which is forbidden for the complete anyon theory of a strictly two-dimensional bosonic gapped ground state but is consistent for a mixed density matrix~\cite{sohal2025noisy}.
This fixed point is consequently intrinsically mixed-state topological order.
We identify the finite-$p$ ordered branch with the same imTO phase provided it is connected to the fixed point without an intervening transition. 

This construction can be stated precisely as higher-form SW-SSB.
For a closed Wilson loop $W_a(\Gamma)$, strong invariance means that its left and right actions stabilize the density matrix separately, whereas weak invariance requires only the diagonal action
$W_a(\Gamma)\rho W_a^\dagger(\Gamma)=\rho$.
Incoherently proliferating short $\psi$ strings preserves a one-sided loop only when its monodromy with $\psi$ is trivial.
Since $M_{\psi,\psi}=+1$, the $\psi$ Wilson-loop symmetry remains strong. 
Here $M_{a,b}$ denotes the monodromy between anyon $a$ and $b$. 
By contrast, $M_{\sigma,\psi}=-1$: a one-sided $\sigma$ loop detects the proliferated $\psi$ strings and is no longer a strong symmetry, but the signs cancel for its simultaneous ket--bra action.
The non-invertible one-form symmetry generated by the $\sigma$ line is therefore reduced from strong to weak, while the surviving strong sector is $\{1,\psi\}$.
Because the surviving fermion is transparent, that sector is nonmodular, giving the one-form-symmetry characterization of the imTO fixed point~\cite{lessa2025strong,sohal2025noisy,zhang2025strong}.

\section{Gapless Majorana at the temporal interface}

\subsection{Bloch Hamiltonian and Chern phases}

We set $J_x=J_y=J>0$ and use primitive vectors
$\bm a_1=(1/2,\sqrt3/2)$ and
$\bm a_2=(1/2,-\sqrt3/2)$.
With $k_a=\bm k\cdot\bm a_a$, the uniform-gauge Majorana Hamiltonian has the two-sublattice form for the left copy (the right copy is the complex conjugate)
\begin{equation}
 h(\bm k)=d_x(\bm k)\tau^x+d_y(\bm k)\tau^y+d_z(\bm k)\tau^z,
 \label{eqS:bloch}
\end{equation}
where $\tau$ acts on sublattices and
\begin{align}
 d_x-\ii d_y=\ii\left(J_z+Je^{-\ii k_1}+Je^{\ii k_2}\right), \qquad
 d_z=\kappa\left[-\sin k_1-\sin k_2+\sin(k_1+k_2)\right].
 \label{eqS:dz-k}
\end{align}
We can analyze the gap closing point and deduce the Chern number.
At $\Gamma=(0,0)$ and $M_1=(\pi,0)$, $M_2=(0,\pi)$, $M_3=(\pi,-\pi)$, the time-reversal-breaking term vanishes.
The four real masses are
\begin{equation}
    m_\Gamma=J_z+2J, \qquad m_1=m_2=J_z, \qquad m_3=J_z-2J.
\label{eqS:four-masses}
\end{equation}
Their signs determine the Chern number of the occupied matter-Majorana band, as summarized in Table~\ref{tab:chern}.

\begin{table}[h]
 \centering
 \caption{Mass signs and Chern number for $J_x=J_y=J>0$ and $\kappa>0$.}
 \label{tab:chern}
 \begin{tabular}{c c c}
 \hline\hline
 range of $J_z$ & $(m_\Gamma,m_1,m_2,m_3)$ & $C$\\
 \hline
 $J_z<-2J$ & $(-,-,-,-)$ & $0$\\
 $-2J<J_z<0$ & $(+,-,-,-)$ & $-1$\\
 $0<J_z<2J$ & $(+,+,+,-)$ & $+1$\\
 $J_z>2J$ & $(+,+,+,+)$ & $0$\\
 \hline\hline
 \end{tabular}
\end{table}

The gap closes at $\Gamma$ when $J_z=-2J$, at $M_1$ and $M_2$ when $J_z=0$, and at $M_3$ when $J_z=2J$.
In particular, the selected strange-correlator geometry uses
\begin{equation}
 0<J_z^{\rm top}<2J,
 \qquad J_z^{\rm ref}>2J,
 \label{eqS:selected-interface}
\end{equation}
so only $m_3$ changes sign.
The topological mismatch is $\Delta C=1$ and the temporal interface carries one Majorana mode at $M_3$.

For completeness, the other possible trivial references expose different sets of modes:
\begin{align}
 -2J<J_z^{\rm top}<0,\ J_z^{\rm ref}<-2J &: \quad \Gamma,\nonumber\\
 -2J<J_z^{\rm top}<0,\ J_z^{\rm ref}>2J &: \quad M_1,M_2,M_3,\nonumber\\
 0<J_z^{\rm top}<2J,\ J_z^{\rm ref}<-2J &: \quad \Gamma,M_1,M_2.
 \label{eqS:other-interfaces}
\end{align}
The single-$M_3$ choice avoids additional flavors and the anisotropic multicritical mode at $\Gamma$.

\subsection{Long-wavelength modes}

Expanding Eq.~\eqref{eqS:bloch} about the $M_3$ point gives,
\begin{align}
 h_{M_3}(\bm q)&=-\sqrt3Jq_y\tau^x+2\kappa q_x\tau^z-m_3 \tau^y.
 \label{eqS:cone-expansions}
\end{align}
with velocities $v_x=2\kappa$ and $v_y=\sqrt3J$.
We now derive the temporal-interface correlator and the quadratic action quoted in the main text.
Combine the two matter Majoranas into
$f_{\bm q}=(c_{A,\bm q}+\ii c_{B,-\bm q})/2$ and use the Nambu spinor
$\Psi_{\bm q}=(f_{\bm q},f_{-\bm q}^\dagger)^T$.
The singular $M_3$ block can be written as
\begin{equation}
 h_{\rm BdG}(\bm q)= 
 \begin{pmatrix}
  \varepsilon_{\bm q} & \Delta_{\bm q}\\
  \Delta_{\bm q}^* & -\varepsilon_{\bm q}
 \end{pmatrix},
 \qquad \varepsilon_{\bm q}=m_3+O(q^2), \qquad \Delta_{\bm q}=v_xq_x-\ii v_yq_y+O(q^2),
 \label{eqS:bdg-block}
\end{equation}
 

The negative-energy eigenmode gives the BCS form
\begin{equation}
 \lvert\Psi\rangle=\mathcal N^{-1}
 \prod_{\bm q>0}\left(1-g_{\bm q}f_{\bm q}^\dagger f_{-\bm q}^\dagger\right)\lvert0\rangle,
 \qquad
 g_{\bm q}=\frac{\xi_{\bm q}-\varepsilon_{\bm q}}{\Delta_{\bm q}^*}, \qquad \xi_{\bm q}=(\varepsilon_{\bm q}^2+|\Delta_{\bm q}|^2)^{1/2}
 \label{eqS:bcs-state}
\end{equation}
In the topological state $m_3<0$, the numerator approaches a nonzero constant as $\bm q\to0$; the trivial state with $m_3>0$ may be adiabatically flattened to the vacuum without changing the singular part of the transition amplitude.
Therefore, a straightforward calculation yields, 
\begin{align}
 \frac{\langle\Psi_{\rm ref}\rvert f_{\bm q,L}f_{-\bm q,L}\lvert\Psi_{\rm top}\rangle}{\langle\Psi_{\rm ref}\vert\Psi_{\rm top}\rangle} &\sim\frac{1}{v_xq_x+\ii v_yq_y},\nonumber\\
 \frac{\langle\Psi_{\rm ref}^*\rvert f_{\bm q,R}f_{-\bm q,R}\lvert\Psi_{\rm top}^*\rangle}
 {\langle\Psi_{\rm ref}^*\vert\Psi_{\rm top}^*\rangle}
 &\sim\frac{1}{-v_xq_x+\ii v_yq_y}.
 \label{eqS:interface-correlators}
\end{align}
where in the second line, we use the time-reversal conjugate to get the right copy. 

These are precisely the diagonal propagators of the two-component real field
$\chi=(\chi_L,\chi_R)^T$ with action
\begin{equation}
 S_M[\chi]=\frac12\int\frac{d^2q}{(2\pi)^2}\,
 \chi^T(-\bm q)
 \left(\ii v_yq_y\mathbbm1+v_xq_x\sigma^z\right)\chi(\bm q).
 \label{eqS:majorana-action}
\end{equation}
Here $\sigma^z$ acts on the copy/chirality index, whereas $\tau^{x,y,z}$ in Eq.~\eqref{eqS:bloch} acts on the microscopic sublattice spinor.
$\chi_L$ and $\chi_R$ are the two chiral components exchanged by the time-reversal-like copy symmetry; together they constitute one nonchiral Majorana field, namely the $c=1/2$ Ising CFT. 


\section{Tricritical-Ising endpoint and supersymmetric exponents}


The most general local continuum action needed near the single-$M_3$ interface is
\begin{equation}
 S[\chi,\phi]=S_M[\chi]+\int d^2x\left[
 \frac12(\partial\phi)^2+\frac{r}2\phi^2 +\frac{u}{4!}\phi^4+g\phi\,\ii\chi_R\chi_L\right].
 \label{eqS:yukawa}
\end{equation}
Here \(\phi\) is a real scalar corresponding to the inter-copy fermion bilinear. Through the Jordan--Wigner equivalence, the massless Majorana theory is the critical Ising model, which admits a bosonic description in terms of the order-parameter field \(\sigma\). The Majorana bilinear \(i\chi_R\chi_L\) is the fermion mass operator and maps to the Ising thermal operator \(\epsilon\), represented by \(\sigma^2\) in Landau--Ginzburg theory:
\begin{equation}
 \ii\chi_R\chi_L \sim \epsilon \sim  \sigma^2.
 \label{eqS:jw-energy}
\end{equation}
A Landau--Ginzburg representation that keeps both fields is therefore
\begin{align}
 S[\sigma,\phi]=\int d^2x\bigg[&\frac12(\partial\sigma)^2 +\frac{r_\sigma}{2}\sigma^2+\frac{u_\sigma}{4}\sigma^4 +\frac{w_\sigma}{6}\sigma^6+g\phi\sigma^2+\frac12(\partial\phi)^2+\frac{r}2\phi^2 +\frac{u}{4!}\phi^4\bigg].
 \label{eqS:two-field-lg}
\end{align}

The Chern-number mismatch supplies one chiral interface branch in each copy.
The mass $\ii\chi_R\chi_L$ is nevertheless forbidden in the symmetric phase by the independent copy parities $\mathbb Z_{2,L}^f\times\mathbb Z_{2,R}^f$.
Because changing $p$ preserves these parities, the dephasing trajectory obeys
\begin{equation}
 r_\sigma=0
 \label{eqS:massless-trajectory}
\end{equation}
until the inter-copy field condenses and breaks them to diagonal parity.
Integrating out the bare-massive component of $\phi$ gives, to leading order,
\begin{equation}
    S_{\rm eff} = \int d^2x\bigg[\frac12(\partial\sigma)^2 +\frac{u_{\sigma}^{\rm eff}}{4}\sigma^4 +\frac{w_\sigma}{6}\sigma^6 \bigg], \qquad u_\sigma^{\rm eff}=u_\sigma-\frac{2g^2}{r}\,.
\end{equation}
Thus dephasing tunes the interaction within the massless manifold.
The three regimes are
\begin{equation}
 \begin{array}{ccl}
 u_\sigma^{\rm eff}>0 &:& \text{Majorana/Ising critical phase},\quad c=1/2,\\
 u_\sigma^{\rm eff}=0 &:& \text{tricritical-Ising endpoint},\quad c=7/10,\\
 u_\sigma^{\rm eff}<0 &:& \text{massive phase}.
 \end{array}
 \label{eqS:three-regimes}
\end{equation}

The tricritical-Ising CFT is the $\mathcal N=1$ supersymmetric minimal model with central charge $c=7/10$.
The multiplet relevant here contains the bosonic primary $\epsilon$, its fermionic descendant $\psi_\epsilon$, and the bosonic superdescendant $\epsilon'$:
\begin{equation}
 x_\epsilon=\frac15,
 \qquad x_{\psi_\epsilon}=\frac7{10},
 \qquad x_{\epsilon'}=\frac65.
 \label{eqS:susy-multiplet}
\end{equation}
Their dimensions differ successively by $1/2$, as required by the supercharges~\cite{friedan1985superconformal,rahmani2015emergent,obrien2018selfdual}.
The microscopic inter-copy boson overlaps with $\epsilon$, while the interface Majorana overlaps with $\psi_\epsilon$.
In two Euclidean dimensions this gives
\begin{equation}
 \avg{\phi(r)\phi(0)}\sim r^{-2x_\epsilon}=r^{-2/5},
 \qquad
 \avg{\chi(r)\chi(0)}\sim r^{-2x_{\psi_\epsilon}}=r^{-7/5}.
 \label{eqS:susy-correlators}
\end{equation}
With the conventional definitions
$\avg{\phi(r)\phi(0)}\sim r^{-\eta_b}$ and
$\avg{\chi(r)\chi(0)}\sim r^{-(1+\eta_f)}$, Eq.~\eqref{eqS:susy-correlators} yields $\eta_b=\eta_f=\frac25$.

Changing the dephasing probability preserves Eq.~\eqref{eqS:massless-trajectory} and therefore couples to $\epsilon'$, not to $\epsilon$.
Its renormalization-group eigenvalue and correlation-length exponent are
\begin{equation}
 y_p=2-x_{\epsilon'}=\frac45,
 \qquad \nu_p=\frac1{y_p}=\frac54.
 \label{eqS:nu-parallel}
\end{equation}
A generic perturbation perpendicular to the topological interface trajectory directly changes the Majorana mass and couples to $\epsilon$ instead:
\begin{equation}
 y_\perp=2-x_\epsilon=\frac95,
 \qquad \nu_\perp=\frac59.
 \label{eqS:nu-perp}
\end{equation}
The distinction between Eqs.~\eqref{eqS:nu-parallel} and \eqref{eqS:nu-perp} is essential: $5/4$ is a property of the topology-constrained dephasing path, not the generic thermal exponent of tricritical Ising.

\end{widetext}

\end{document}